\documentclass[prb,aps,twocolumn,showpacs,amsmath,amssymb,superscriptaddress,floatfix,reprint]{revtex4-1}
\usepackage{graphicx}
\usepackage{dcolumn}
\usepackage{bm}
\usepackage{epsfig}

\begin{document}

\title{Odd-frequency pairing in (TMTSF)$_2$ClO$_4$}

\author{F.L. Pratt}
\affiliation{ISIS Facility, STFC Rutherford Appleton Laboratory, Chilton, Oxfordshire OX11 0QX,UK}

\author{T. Lancaster}
\affiliation{Centre for Materials Physics, Department of Physics, Durham University, South Road, Durham DH1 3LE, UK}

\author{S.J. Blundell}
\affiliation{University of Oxford, Department of Physics, Clarendon Laboratory, Parks Road, Oxford OX1 3PU, UK}

\author{C. Baines}
\affiliation{Swiss Muon Source, Paul Scherrer Institut, Villigen, CH-5232, Switzerland}

\date{\today}

\begin{abstract} 
The low field phase of the organic superconductor (TMTSF)$_{2}$ClO$_{4}$ is studied by muon spin rotation. 
The zero temperature limit of the magnetic penetration depth within the TMTSF layers is obtained to be $\lambda_{ab}(0) = 0.86(2)~\mu$m.
Temperature dependence of the muon spin relaxation shows no indication of gap nodes on the Fermi surface nor of any spontaneous fields due to time-reversal-symmetry breaking. 
The weight of evidence suggests that the symmetry of this low field phase is odd-frequency  $p$-wave singlet; a novel example of odd-frequency pairing in a bulk superconductor.
\end{abstract}

\maketitle

The Bechgaard salts\cite{Jerome} (TMTSF)$_{2}$X (anion X typically being PF$_6$ or ClO$_4$) attract continuing interest
as a system whose rich physics is derived from a quasi-one-dimensional character and strong electron-electron interactions \cite{IshiguroYamaji,Toyota07,lebed2}. 
Although (TMTSF)$_{2}$ClO$_4$ (TMC) was the first ambient pressure organic superconductor (SC) to be discovered\cite{Bechgaard81}, 
the exact nature of its unconventional SC phases has yet to be resolved,
with recent studies indicating that there is more than one distinct SC phase in this system.
One phase is observed up to ${\mu}_0H_{c2}$ = 0.16 T for magnetic fields $H$ oriented perpendicular to the TMTSF layers\cite{IshiguroYamaji,yonezawa} (low-field L-SC phase), 
but aligning $H$ within the TMTSF layers increases $H_{c2}$ significantly above the Pauli limit \cite{OhNaughton04} and allows a further high-field SC phase to be revealed. 
NMR finds clear evidence\cite{Shinagawa07} for a field-induced transition between the L-SC phase and the second (H-SC) phase for fields above $\sim$1.5 T;
the L-SC phase shows the Knight shift expected for a singlet state, whereas the H-SC does not, indicating a triplet state. 

A fundamental parameter for a superconductor is the magnetic penetration depth $\lambda$, which determines the superfluid stiffness $\rho = c^2/ \lambda^2$ characterising the electromagnetic response to an applied magnetic field.
The variation of $\rho(T)$ or $\lambda(T)$ with temperature $T$ provides an important test of the gap symmetry.
Previous estimates of $\lambda_{ab}(0)$ for TMC using
muon-spin rotation ($\mu$SR) have either given a value around 1.2~$\mu$m \cite{HarshmanMills92,Le93,Greer03} 
or a distinctly smaller value in the region of 0.5~$\mu$m \cite{Luke03}. 
None of these studies took into account the strong $H$ variation of the line width due to the small value of $H_{c2}$ when the supercurrents are in the $ab$ plane.
In order to resolve this $\lambda_{ab}(0)$ discrepancy and gain further information about the SC gap symmetry, we have made a detailed $\mu$SR study of this system, 
investigating the dependence of the SC properties over a full range of $H$ and $T$.

From these results and other previously reported properties, we find strong evidence that the symmetry of the L-SC phase in TMC is an unusual odd-frequency pairing odd-parity singlet state. 
Odd frequency pairing \cite{Tanaka12} was first proposed\cite{Berezinsky74} for $^3$He and subsequently for superconductors\cite{BalatskyAbrahams92}.
It is generally believed to be most relevant to symmetry-broken local environments, such as those near interfaces or in vortex cores \cite{Tanaka12} and has not previously been verified in a bulk superconductor. 

Crystals of TMC form as needles aligned with the molecular stacking direction $a$. 
This is the most conducting direction of the material, which has a triclinic structure in
which layers of TMTSF are arranged in the $ab$ plane, separated along the c axis by layers of the ClO$_4$ anions. 
The $c^*$  direction perpendicular to the molecular planes is the axis with the smallest conductivity. 
Our sample consisted of a  mosaic of these crystals (total mass 124~mg) aligned with their $a$ axes parallel. 
In the measurements $H$ is applied perpendicular to $a$, so we measure an axial average of the properties.

Transverse field muon spin rotation (TF $\mu$SR) provides a means of
accurately measuring the internal magnetic field distribution caused by
the vortex lattice (VL) in a type II SC \cite{sonier}
and has previously proven useful in establishing the vortex properties of molecular SCs
\cite{pratt01,lancaster}.
Our field-cooled TF $\mu$SR measurements used two $\mu^+$SR instruments.
At the Swiss muon source (S$\mu$S) within the Paul Scherrer Institute we used the LTF dilution refrigerator instrument to measure down to 20 mK.
At the ISIS Facility, STFC Rutherford Appleton Laboratory, UK we used a $^3$He sorption cryostat with the MuSR instrument to measure down to 300 mK. 
For the LTF measurements, magnetic field $B_{\mathrm{app}} = \mu_0H$ 
was applied perpendicular to both $a$ and the plane of the sample. 
For the MuSR measurements, $B_{\mathrm{app}}$ was perpendicular to $a$ and parallel to the plane of the sample mosaic.
To ensure that the sample was in the non-magnetic fully relaxed anion-ordered (AO) state
the cooling rate was maintained at 1~K h$^{-1}$ to 2~K h$^{-1}$ between 35 K and 15 K. 
All data analysis was carried out using the WiMDA program \cite{WiMDA}.

In TF $\mu$SR measurements performed above $T_{\mathrm{c}}$, the distribution of field $B$ at the muon,  $p(B)$,  is broadened mainly by randomly
oriented nuclear moments near the muon stopping sites; in this case dominated by the methyl groups at the ends of the TMTSF molecules.
This leads to an essentially Gaussian relaxation of the muon polarisation 
$P_{x}(t) \propto e^{-\sigma_{\mathrm{n}}^{2}t^{2}/2} \cos (\gamma_{\mu} \langle B\rangle t + \phi)$, 
where $\sigma_{\mathrm{n}}^{2} = \gamma^{2}_{\mu}\langle (B - \langle B\rangle)^{2} \rangle$.
For  $B_{\mathrm{app}}=2.5$~mT the TF broadening $\sigma_{\mathrm{n}}/\gamma_{\mu}$ was monitored 
on cooling through the AO region and the width was found to reduce by about 10\% on ordering.
The AO doubles the $b$ axis leading to two inequivalent TMTSF stacks and significant distortion of the methyl groups on one of the stacks leading to a non planar conformation\cite{Pevelen01}. 
The observed magnitude of change in $\sigma_{\mathrm{n}}$ is consistent with this distortion.  

\begin{figure}[tb]
\begin{center}
\epsfig{file=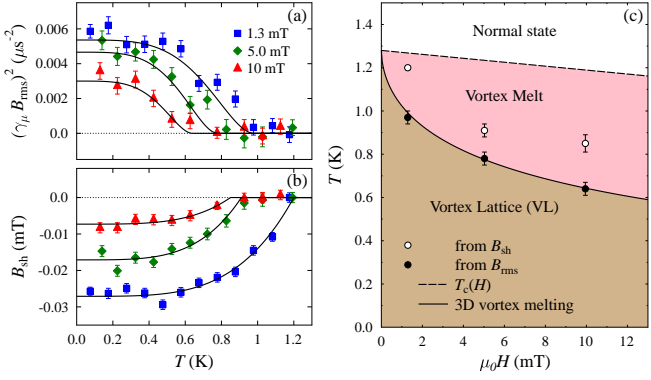,width=8.7cm}
\caption{(a) SC line width $B_{\mathrm{rms}}$ 
and (b) average field shift $B_{\mathrm{sh}}$ versus $T$ for several different values of applied field measured using LTF. 
Characteristic transition temperatures $T'$ are obtained from two-fluid model fits (lines).
(c) The low field phase diagram, comparing $T'$ from $\mu$SR (points) with previously reported\cite{yonezawa} $T_{\mathrm{c}}(H)$ behaviour for the $H{\parallel}c^*$ orientation that dominates here (dashed line). 
$T'$ from $B_{\mathrm{rms}}$ is consistent with a 3D vortex melting curve (solid line)\cite{Brandt95}.
\label{vortexfig}}
\end{center}
\vspace{-0.7cm}
\end{figure}

\begin{figure}[tb]
\begin{center}
\epsfig{file=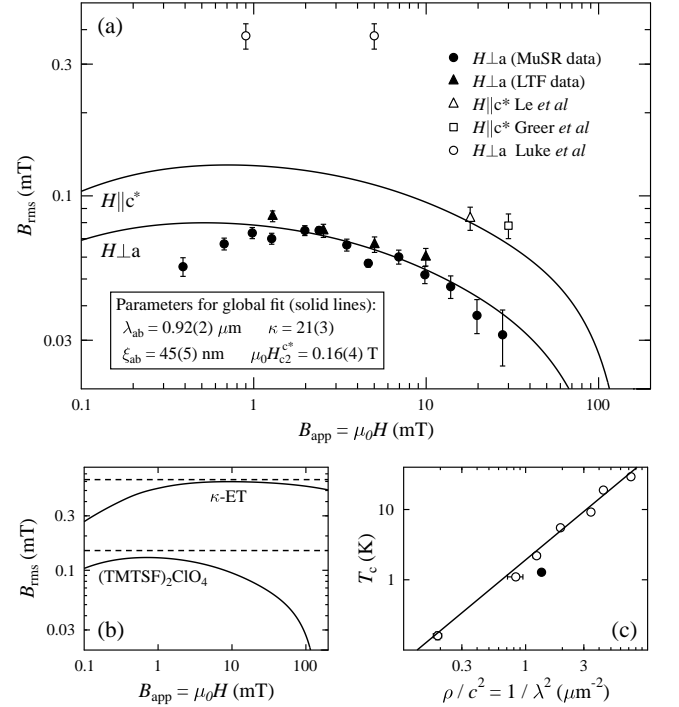,width=8.7cm}
\vspace{-0.3cm}
\caption { (a) Field dependent SC line width at 0.35~K (solid points).
Open points are data taken from Refs.~\onlinecite{Le93,Greer03,Luke03}. 
The global fit is described in the text (solid lines). 
(b) Comparison with $B_{\mathrm{rms}}(H)$ for the typical $\kappa$-ET organic superconductor.
Dashed lines show the corresponding Barford-Gunn\cite{BarfordGunn} field-independent values.
(c) Scaling plot of $T_c$ against $\rho(0)$ for molecular SCs. 
The filled point is the present result, open points are previous data for molecular SCs\cite{PrattBlundell}.}
\label{GLfig}
\end{center}
\vspace{-1cm}
\end{figure}

In the SC state the VL contributes a further broadening, giving $\sigma ^2= \sigma^2_{\mathrm{n}} + \sigma^2_{\mathrm{VL}}$
with corresponding VL field width $B_{\mathrm{rms}} = \sigma_{\mathrm{VL}}/\gamma_{\mu}$. 
Due to the relatively weak VL contribution,  we did not attempt to spectrally resolve the 
asymmetric VL field profile as we did in previous studies of organic SC such as
$\kappa$-(ET)$_2$Cu(NCS)$_2$ \cite{Lee97,Pratt05} ($\kappa$-ET).
The $T$ dependence of $B_{\mathrm{rms}}$ 
and the shift of the average internal field $B_{\mathrm{sh}}=\langle B\rangle-B_{\mathrm{app}}$ 
at several
values of $B_{\mathrm{app}}$
are shown in Fig.\ref{vortexfig} for data measured using LTF.
Consistent data are obtained between runs on LTF and MuSR despite the difference in 
orientation of $B_{\mathrm{app}}$, indicating 
that the mosaic sample provides a good axial average about the $a$ axis.  
Both $B_{\mathrm{rms}}$ and $|B_{\mathrm{sh}}|$ 
fall continuously with increasing $T$, reaching zero above a characteristic temperature $T'$.
The effect of increasing  $B_{\mathrm{app}}$ is clearly seen as a depression of both $T'$ and the 
size of the SC response measured by $B_{\mathrm{rms}}$ and $|B_{\mathrm{sh}}|$ at low $T$.
In order to extract characteristic $T'$ values, fits were made for $T< T'$ to a simple two-fluid $T$ dependence, i.e. $1- t^4$, where $t = T/T'$. 
The obtained values are shown in Fig.\ref{vortexfig}c for 
different values of $H$. 
There is an offset of around 0.2~K between values of $T'$ obtained from  $B_{\mathrm{rms}}$ and from $B_{\mathrm{sh}}$. 
Previous transport studies\cite{Yonezawa09} suggested a vortex liquid  phase just below $T_{\mathrm{c}}$; we find that 
$T'$ obtained from $B_{\mathrm{rms}}$ is well described by a 3D vortex melting curve
and we assign this $T'$ to the melting  temperature $T_{\mathrm{m}}$.
The melting curve at low $H$ takes the form \cite{Brandt95} $T_{\mathrm{m}}(H)=T_{\mathrm{c}}(0)/(1+(H/H_0)^{1/2})$ where $H_0$ is a characteristic field.  
In highly 2D systems such as $\kappa$-ET, field-induced layer decoupling transitions dominate over 3D melting \cite{Pratt05,lancaster}, 
but strong Josephson coupling stabilises the layers for TMC.

Having established the stability of a VL phase over a range of $T$ and $B_{\mathrm{app}}$, we focus on the accurate determination of $\lambda$. 
A reliable $\lambda$ value can be obtained by measuring $B_{\mathrm{rms}}(B_{\mathrm{app}})$ at low $T$ and comparing with the prediction for an ideal vortex lattice calculated using the Ginzburg-Landau (GL) model \cite{brandt}. 
The expected behaviour is an increase of $B_{\mathrm{rms}}$ with increasing $B_{\mathrm{app}}$ in the region of $B_{\rm{c1}}$, followed by a maximum and then a fall towards  
zero at $B_{\mathrm{c}2}$. 
The measured $B_{\mathrm{rms}}(B_{\mathrm{app}})$
at $T$~=~0.35~K  is shown in Fig.~\ref{GLfig}a, along with the earlier reported results. 
For fitting the data we assume an angular variation for $\xi$ and $\lambda$ of the form $(\cos^2 \theta + \sin^2 \theta/ \gamma^2)^{-1/4}$, 
where $\theta$ is the angle to $c^*$ and the anisotropy factor $\gamma$ is obtained from the $B_{\rm c2}$ slope ratio\cite{yonezawa} as $(2.3/0.11)^{1/2} = 4.5$.
The overall behaviour follows the GL form, consistent with a well defined VL at low $T$. 
Data above 1 mT are fitted by the parameters $\lambda = 0.92(2)~\mu$m and $\kappa$ = 21(3).
The corresponding $B_{c2}$ value of 0.16(4) T is fully consistent with direct measurements \cite{IshiguroYamaji,yonezawa}. 
The results of Refs.~\onlinecite{Greer03,Le93}, which were made on aligned crystals
with $B_{\mathrm{app}}\parallel c^*$ are also consistent with our fit and our analysis includes these data. 
In contrast, the data points of Ref.~\onlinecite{Luke03} have an anomalously large $B_{\mathrm{rms}}$ and are excluded from the global fit. 
The small $B_{\mathrm{c}2}$ for $B_{\mathrm{app}}{\parallel}c^*$ leads to a remarkably strong field dependence of $B_{\mathrm{rms}}$ in our measurement range 
and the extended plateau region typical of strong type-II (large $\kappa$) materials is not present here. 
In contrast, $\kappa$-ET (Fig.2b) has a more well-developed plateau due to its larger values of $\kappa$ and $B_{\mathrm{c}2}$,
along with an enhanced $B_{\mathrm{rms}}$ scale resulting from its smaller $\lambda$.  

\begin{figure}[tb]
\begin{center}
\epsfig{file=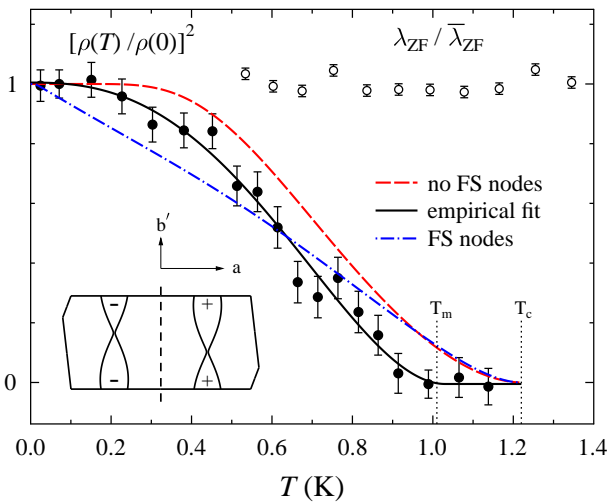,width=8.5cm}
\caption {
TF data at 1.9 mT taken down to 20~mK (solid circles).
The square of $\rho(T)$ normalised to $\rho(0)$ is plotted along with an empirical power law fit (solid line). 
The linear dependence for states with line nodes on the FS (dot-dash line) is not seen in the data, which are much closer to fully gapped BCS behaviour (dashed line).
The inset illustrates the anion-ordered FS in the $ab$ plane calculated by Nagai {\em et al}\cite{Nagai11}; 
our data are consistent with a gap function $\Delta({\bf k})$ having a nodal plane parallel to the warped FS sheets (dashed line), e.g. $\Delta({\bf k})=\Delta_0 \sin k_a$. 
The ZF relaxation rate (open circles) shows no significant change below  $T_{\mathrm{c}}$.
}
\vspace{-0.60cm}
\label{lowT}
\end{center}
\end{figure}

Breaking of time-reversal symmetry (TRS) in some types of unconventional SC can lead to spontaneous internal fields \cite{sigrist} in 
zero applied field (ZF). These weak fields have been observed by $\mu$SR in several examples,
e.g. Sr$_2$RuO$_4$  \cite{luke98} and 
more recently LaNiC$_{2}$ \cite{hillier}.  
The spontaneous fields lead to an increase in the magnitude
of the ZF depolarization rate of the muon polarization in the SC phase.  
To search for this effect, we took ZF $\mu$SR data as a function of $T$, 
scanning through $T_{\mathrm{c}}$. 
Measured spectra were fitted to an exponential relaxation function $e^{-{\lambda}_{\rm{ZF}}t}$ and $\lambda_{\rm{ZF}}$ is shown
in Fig.~\ref{lowT} (open circles).
Within experimental error we observe no evidence for a spontaneous local magnetic field, consistent with an earlier report\cite{Luke03}. 

In order to gain further information on the SC pairing symmetry present in TMC we have examined the TF broadening down to 20 mK,  taking relatively high statistics LTF data in a field of 1.9~mT.
This field was chosen to be close to the maximum of $B_{\mathrm{rms}}$ and a field at which the VL is thermally stable over a wide range of $T$.
The square of the superfluid density $\rho(T)$, normalised to its low $T$ value $\rho(0)$, is shown in Fig.\ref{lowT} (solid points), each point corresponds to $\sim$10$^7$ analysed muon decay events.
The observed reduction of $\rho(T)$ with increasing $T$ reflects the excitation of quasiparticles, which is highly sensitive to the quasiparticle density of states (QDOS) and thus the presence of any gap nodes on the Fermi surface (FS).
A good description of the data is provided by the empirical power law $\rho(T)/\rho(0) = 1- t^n (t=T/T_{\rm m})$.
Our fitted value is $n$ = 2.5(3), which is not consistent with the $n$ = 1 linear behaviour predicted for models with FS line nodes and seen previously, for example, in $\mu$SR studies of $\kappa$-ET\cite{pratt01,Le92}. 
The clear saturation of $\rho(T)$ at low $T$ is strongly indicative of a fully gapped SC with no FS nodes.  
This fully gapped state agrees with a previous thermal conductivity study \cite{BelinBehnia97}, but differs from the conclusions of a recent $ab$ plane angle-resolved heat capacity study\cite{yonezawa},
in which FS gap nodes were inferred \cite{yonezawa,Nagai11} from dip structure found in $C/T$ data at 0.14~K and 0.3~T, when field was oriented at $\pm$10$^{\circ}$ to $a$. 
The only nodal symmetry that would be consistent with our data, however, is a $p_x$ state having no nodes crossing the FS (Fig.3 inset).

\setlength{\tabcolsep}{3.5mm}
\begin{table*}[htb]
\begin{tabular}
{ |    l            l             c                c                        c                        c                              c                                   c                                            c|}
\hline
 Symmetry & Type                  & Broken   & FS nodes$^a$       &  Pauli   &  Knight       & Disorder                         & HS Peak & Label and \\ 
 ($\omega,S,k$)              &       & TRS       &     & limit   &   shift                            & lowers $T_c$               &                             &               Assignment \\
\hline
   &          & {\bf No}\cite{Luke03}$^{,b}$       &{\bf  No}\cite{BelinBehnia97}$^{,b}$ Yes\cite{yonezawa}      &        -              &  Yes\cite{Shinagawa07}                & {Yes}\cite{Coulon82,Joo04}                       & No\cite{HS-peak,Shinagawa07}                     &  L-SC  \\
    &        & ?                                        & No\cite{yonezawa}             &      No\cite{OhNaughton04}             &  No\cite{Shinagawa07}                   & ?                       & No\cite{Shinagawa07}                                    & H-SC \\               
\hline
O S O &$p$-singlet &      No      &        No        &          Yes             &              Yes                  &      Yes                     &      No                       & \bf{L-SC}\\
 & & & & & & & &\\
O T E & $s$-polar &     No       &       No         &         No              &            No                    &   No                    &    No                         & H-SC\\
&$s$-$\beta$ &     Yes        &      No        &        No               &           No                     &   No                   &     No                        &   \\
&$s$-BW &     No       &        No     &          No             &            Possible$^{c}$                   &    No                  &      No                       &   \\
 & & & & & & & & \\
E S E & $s$-singlet   & No      &No              &  Yes                &  Yes                            & No                      & Yes                     &     \\
& & & & & & & & \\
E T O&$p$-polar  &      No      &        No          &            No           &                No                &       Yes                    &      No                       &     H-SC\\
&$p$-$\beta$  &     Yes       &       No        &           No            &               No                 &       Yes                    &      No                       &    \\
&$p$-BW &      No      &         No   &             No          &                 Possible$^{c}$               &        Yes                   &       No                      &   \\
  \hline
\end{tabular}
\caption{
Properties of the superconducting states of TMC, experimental (top two lines) and theoretical (based on Powell\cite{Powell}). 
Bold text marks our main conclusions and unique assignment for L-SC.
(E=even, O=odd, S=singlet, T=triplet, HS=Hebel-Slichter, $^a$experimental result or symmetry requirement from theory, $^b$this study, $^c$direction dependent).
}
\vspace{-0.5cm}
\end{table*}

From our $\rho(T)$ dependence we can extrapolate to $T$~=~0, giving $\lambda$(0)~=~0.86(2)~$\mu$m. 
Scaling behaviour $T_{\mathrm{c}} \propto \rho^{m}$ has been explored for many classes of SC, including cuprates \cite{uemura1,uemura2} and pnictides\cite{clarke}, where $m = 1$, and molecular systems \cite{PrattBlundell} where $m=3/2$. 
The new $\rho$ value places TMC significantly below the molecular SC trend line\cite{PrattBlundell} (Fig.~\ref{GLfig}c).
Introducing non-magnetic impurities to TMC strongly suppresses $T_{\mathrm{c}}$\cite{Coulon82,Joo04}, but regular samples are extremely pure.  
The low $T_{\mathrm{c}}$ therefore suggests a different pairing mechanism from the other molecular SCs. 

In order to systematically assign the SC phases, we summarize in Table 1 key experimental properties, both in the low field L-SC state studied here and also in the high field H-SC state established for in-plane fields above 1.5~T (Table 1, upper section), along with the eight specific theoretical possibilities for a P-1 triclinic system with inversion symmetry, previously identified in the group theory analysis of Powell\cite{Powell} (Table 1, lower section).
For the L-SC phase, it can be seen that the only assignment consistent with experiment is the OSO state (odd-frequency singlet with odd-parity).

Although early theoretical studies suggested that such an odd-frequency SC would be intrinsically gapless \cite{BalatskyAbrahams92},
subsequent work has indicated that gapless behaviour is not an essential feature.
In particular, a model based on a condensate of composite bosons made up of a Cooper pair and a magnon  has recently been explored in some detail\cite{Dahal09}. 
Depending on model parameters, the QDOS for this model could range from the standard sharply-peaked form of the fully-gapped Bardeen-Cooper-Schrieffer (BCS) model through to a form with a completely closed gap.
In an intermediate regime the QDOS peak becomes broadened and weakened compared to BCS.
As $\rho(T)$ falls faster than BCS in the region above 0.2 K (Fig.3), this difference would be consistent with the broadening of the QDOS into the gap region expected for the intermediate regime in this type of model\cite{Dahal09}, 
although we note that thermal vortex motion could also lead to $B_{\mathrm{rms}}$ falling with $T$ in a region just below $T_{\rm m}$, giving an apparent extra fall in $\rho(T)$. 

Turning to the H-SC phase we see that there are two possibilities, either an ETO state (even-frequency triplet with odd-parity) of the polar form, as previously suggested by Powell\cite{Powell}, or alternatively an OTE state (odd-frequency triplet with even-parity), also of polar form. 
For a quasi-1D system such as TMC a field-induced transition from OSO to OTE pairing has been found in theoretical studies based on the extended Hubbard model, due to enhancement of charge fluctuations over spin fluctuations\cite{Aizawa08,Shigeta09,Shigeta11}.
Clear discrimination between the two H-SC candidates could be gained by studying the effect of disorder on $T_{\mathrm c}$ within the H-SC phase, however the studies to date have only measured the  L-SC phase\cite{Coulon82,Joo04}. 

Powell\cite{Powell} has suggested a scenario in which the field-induced transition is due to spin-orbit coupling (SOC). 
In this picture a strong-SOC $p$-BW state\cite{Powell} for the L-SC phase crosses over to a weak-SOC $p$-polar state\cite{Powell} in the H-SC phase. 
However, significant SOC is not expected here at low $T$ due to the underlying inversion symmetry once the TMTSF torsional modes have been frozen out \cite{Adrian86}.

Finally, we note that the thermodynamic stability of odd-frequency bulk SCs was initially questioned,
but recent theoretical work now concludes that these states should actually be stable\cite{Solenov09,Kusunose11}. 
The example identified here would suggest that such an odd-frequency state can indeed exist in a bulk material.

We thank N. Toyota for providing the sample, J. Quintanilla and M. Eschrig for discussion and comment and  EPSRC, UK for financial support.
We acknowledge the late E.H. Brandt for providing the numerical data to enable fitting to the GL model. 
Parts of this work were carried out at the STFC ISIS Facility, UK and at the Swiss Muon Source, Paul Scherrer Institute, CH.


\begin{thebibliography}{xx}

\bibitem{Jerome}
D. J\'{e}rome, A. Mazaud, M. Ribault and K. Bechgaard, J. Physique Lett. {\bf 41}, L95 (1980).

\bibitem{IshiguroYamaji}
T. Ishiguro and K. Yamaji and G. Saito {\it Organic Superconductors} (Springer, Berlin, 1998) 2nd ed.

\bibitem{Toyota07}
N.~Toyota, M.~Lang and J.~M\"{u}ller {\it Low-Dimensional Molecular Metals} (Springer, Berlin, 2007).

\bibitem{lebed2}
{\it The Physics of Organic Superconductors and Conductors,} edited by A.G. Lebed (Springer, Berlin, 2008).

\bibitem{Bechgaard81}
K. Bechgaard, K. Carneiro, M. Olsen, F.B. Rasmussen, C.S. Jacobsen, Phys. Rev. Lett. {\bf 46}, 852 (1981).

\bibitem{yonezawa}
S. Yonezawa, Y. Maeno, K. Bechgaard and D. J\'{e}rome,
Phys. Rev. B {\bf 85}, 140502(R) (2012). 

\bibitem{OhNaughton04}
J.I. Oh and M.J. Naughton, 
Phys. Rev. Lett. {\bf 92}, 067001 (2004). 

\bibitem{Shinagawa07}
J. Shinagawa, Y. Kurosaki, F. Zhang, C. Parker, S.E. Brown, D. J\'{e}rome, J.B. Christensen and K. Bechgaard, Phys. Rev. Lett. {\bf 98}, 147002 (2007).

\bibitem{HarshmanMills92}
D.R. Harshman and A.P. Mills Jr., Phys. Rev. B {\bf 45}, 10684 (1992).

\bibitem{Le93}
L.P. Le, A. Keren, G.M. Luke, B.J. Sternlieb, W.D. Wu, Y.J. Uemura, J.H. Brewer, T.M. Riseman, 
R.V. Upasani, L.Y. Chiang, W. Kang, P.M. Chaikin, T. Csiba and G. Gr\"{u}ner,
Phys. Rev. B {\bf 48}, 7284 (1993).

\bibitem{Greer03} 
A.J. Greer, D.R. Harshman, W.J. Kossler, A. Goonewardene, D. Ll. Williams, E. Koster, W. Kang, R.N. Kleiman and R.C. Haddon,
Physica C {\bf 400}, 59 (2003).

\bibitem{Luke03}
G.M. Luke, M.T. Rovers, A. Fukaya, I.M. Gat, M.I. Larkin, A. Savici, Y.J. Uemura, K.M. Kojima, P.M. Chaikin, I.J. Lee and M.J. Naughton,
Physica B {\bf 326}, 378 (2003).

\bibitem{Tanaka12}
Y. Tanaka, M. Sato and N. Nagaosa, J. Phys. Soc. Jpn. {\bf 81}, 011013 (2012).

\bibitem{Berezinsky74}
V. L. Berezinskii, Pis'ma Zh. Eksp. Teor. Fiz. {\bf 20}, 628 (1974) [JETP Lett. {\bf 20}, 287 (1974)].

\bibitem{BalatskyAbrahams92}
A. Balatsky and E. Abrahams, Phys. Rev. B {\bf 45}, 13125 (1992).

\bibitem{sonier}
J.E. Sonier, J.H. Brewer, and R.F. Kiefl, Rev. Mod. Phys. {\bf 72}, 769 (2000).

\bibitem{pratt01}
F.L. Pratt, S.L. Lee, C.M. Aegerter, C. Ager, S.H. Lloyd, S.J. Blundell, F.Y. Ogrin, E.M. Forgan, H. Keller, W. Hayes, T. Sasaki, N. Toyota and S. Endo, 
Synth. Met. {\bf 120}, 1015 (2001).

\bibitem{lancaster}
T. Lancaster, S.J. Blundell, F.L. Pratt, and J.A. Schlueter
Phys. Rev. B {\bf 83}, 024504 (2011). 

\bibitem{WiMDA}
F.L. Pratt, Physica B {\bf 289}, 710 (2000).

\bibitem{Pevelen01}
D. Le P\'{e}velen, J. Gaultier, Y. Barrans, D. Chasseau, F. Castet and L. Ducasse, Eur. Phys. J. B {\bf 19}, 363 (2001).

\bibitem{Lee97} S.L. Lee, F.L. Pratt, S.J. Blundell, C.M. Aegerter, P.A. Pattenden, 
K.H. Chow, E.M. Forgan, T. Sasaki, W. Hayes and H. Keller, Phys. Rev. Lett. {\bf 79}, 1563 (1997).

\bibitem{Pratt05}
F.L. Pratt, S.J. Blundell, T. Lancaster, M.L. Brooks, S.L. Lee, N. Toyota and T. Sasaki,
Synth. Met.  {\bf 152} 417 (2005).

\bibitem{Yonezawa09}
S. Yonezawa, S. Kusaba, Y. Maeno, P. Auban-Senzier, C. Pasquier and D. J\'{e}rome, J. Phys.:Conf. Ser. {\bf 150}, 052289 (2009).

\bibitem{Brandt95}
E.H. Brandt, Rep. Prog. Phys. {\bf 58}, 1465 (1995).

\bibitem{brandt}
E.H. Brandt, Phys. Rev. B {\bf 68}, 054506 (2003).

\bibitem{BarfordGunn}
W. Barford and M. Gunn, Physica {\bf 156C}, 515 (1988). 

\bibitem{sigrist}
M. Sigrist and K. Ueda, Rev. Mod. Phys. {\bf 63}, 239–311 (1991). 

\bibitem{luke98}
G.M. Luke, Y. Fudamoto, K.M. Kojima, M.I. Larkin, M. I. Larkin, J. Merrin, B. Nachumi, Y. J. Uemura, Y. Maeno, Z. Q. Mao, Y. Mori, H. Nakamura and M. Sigrist, 
Nature {\bf 394} 561 (1998).

\bibitem{hillier}
A. D. Hillier, J. Quintanilla, and R. Cywinski, 
Phys. Rev. Lett. {\bf 102}, 117007 (2009).

\bibitem{Le92}
L.P. Le, G.M. Luke, B.J. Sternlieb {\em et al}, Phys. Rev. Lett. {\bf 68}, 1923 (1992).

\bibitem{BelinBehnia97}
S. Belin and K. Behnia,  Phys. Rev. Lett. {\bf 79}, 2125 (1997).

\bibitem{Nagai11}
Y. Nagai, H. Nakamura and M. Machida, Phys. Rev. B {\bf 83}, 104523 (2011).

\bibitem{uemura1}
Y.J. Uemura, G. M. Luke, B. J. Sternlieb, J. H. Brewer, J. F. Carolan, W. N. Hardy, R. Kadono, J. R. Kempton, R. F. Kiefl, S. R. Kreitzman, P. Mulhern, T. M. Riseman, D. Ll. Williams, B. X. Yang, S. Uchida, H. Takagi, 
J. Gopalakrishnan, A. W. Sleight, M. A. Subramanian, C. L. Chien, M. Z. Cieplak, Gang Xiao, V. Y. Lee, B. W. Statt, C. E. Stronach, W. J. Kossler, and X. H. Yu,  
Phys. Rev. Lett. {\bf 62}, 2317 (1989). 

\bibitem{uemura2}
Y. J. Uemura, L. P. Le, G. M. Luke, B. J. Sternlieb, W. D. Wu, J. H. Brewer, T. M. Riseman, C. L. Seaman, M. B. Maple, M. Ishikawa, D. G. Hinks, J. D. Jorgensen, G. Saito, and H. Yamochi, 
Phys. Rev. Lett. {\bf 66}, 2665 (1991).

\bibitem{clarke}
M.J. Pitcher, T. Lancaster, J.D. Wright, I. Franke, A.J. Steele, P.J. Baker, F.L. Pratt, W. T. Thomas, D.R. Parker, S.J. Blundell and  S.J. Clarke,
J. Am. Chem. Soc. {\bf 132}, 10467 (2010).

\bibitem{PrattBlundell}
F.L. Pratt and S.J. Blundell, Phys. Rev. Lett. {\bf 94}, 097006 (2005).

\bibitem{Coulon82}
C. Coulon, P. Delhaes, J. Amiell, J.P. Manceau, J.M. Fabre and L. Giral, J. Physique {\bf 43}, 1721 (1982).

\bibitem{Joo04}
N. Joo, P. Auban-Senzier, C.R. Pasquier, P. Monod {\em et al}, Eur. Phys. J. B {\bf 40}, 43 (2004).

\bibitem{Powell}
B.J. Powell, J. Phys.:Condens. Matter {\bf 20}, 345234 (2008).

\bibitem{Dahal09}
H.P. Dahal, E. Abrahams, D. Mozyrsky, Y. Tanaka and A.V. Balatsky, New J. Phys. {\bf 11}, 065005 (2009).

\bibitem{HS-peak}
M. Takigawa, H. Yasuoka and G. Saito, J. Phys. Soc. Jpn. {\bf 56}, 873 (1987). 

\bibitem{Aizawa08}
H. Aizawa, K. Kuroki and Y. Tanaka, Phys. Rev. B {\bf 77}, 144513 (2008).

\bibitem{Shigeta09}
K. Shigeta, S. Onari, K. Yada, and Y. Tanaka, Phys. Rev. B {\bf  79}, 174507 (2009).

\bibitem{Shigeta11}
K. Shigeta, Y. Tanaka, K. Kuroki, S. Onari and H. Aizawa, Phys. Rev. B {\bf 83}, 140509(R) (2011).

\bibitem{Adrian86}
F.J. Adrian,  Phys. Rev. B {\bf 33}, 1537 (1986).

\bibitem{Solenov09}
D. Solenov, I. Martin and D. Mozyrsky, Phys. Rev. B {\bf 79}, 132502 (2009).

\bibitem{Kusunose11}
H. Kusunose, Y. Fuseya and K. Miyake, J. Phys. Soc. Jpn. {\bf 80}, 054702 (2011).

\end{thebibliography}
\end{document}